\documentclass[a4paper]{aa}
\usepackage{graphicx}
\begin{document}
\title{The magnetohydrodynamic instability of current-carrying jets} 
\author{A.~Bonanno\inst{1,2}, V.~Urpin\inst{1,3}}
\institute{$^{1)}$ INAF, Osservatorio Astrofisico di Catania,
           Via S.Sofia 78, 95123 Catania, Italy \\
           $^{2)}$ INFN, Sezione di Catania, Via S.Sofia 72,
           95123 Catania, Italy \\
           $^{3)}$ A.F.Ioffe Institute of Physics and Technology and
           Isaac Newton Institute of Chile, Branch in St. Petersburg,
           194021 St. Petersburg, Russia}
\date{\today}

\abstract
{Magnetohydrodynamic instabilities can be responsible for the formation of structures with
various scales in astrophysical jets.}
{We consider the stability properties of jets containing both the 
azimuthal and axial field of subthermal strength. 
A magnetic field with complex topology in jets is suggested by theoretical 
models and is consistent with recent   observations.} 
{Stability is discussed by means  of a linear analysis of the ideal magnetohydrodynamic
equations.} 
{We argue that in azimuthal and axial magnetic fields the jet is always unstable to non-axisymmetric perturbations.
Stabilization  does not occur even if the strengths of these field components are comparable. 
If the axial field is weaker than the azimuthal one, instability occurs for 
perturbations with any azimuthal wave number $m$, and the growth rate reaches 
a saturation value for low values of $m$. If the axial field is stronger 
than the toroidal one, the instability 
shows for perturbations with relatively high $m$. }
{}

\titlerunning{MHD instabilities jets}
\maketitle
\section{Introduction} 

The magnetic field  plays a key role in the formation and propagation 
of astrophysical jets. Polarization measurements provide important information 
about the orientation and the strength  of the magnetic field, which seems to be 
highly organized on a large scale in many jets (see, e.g., Cawthorne et al. 1993, 
Lepp\"{a}nen, Zensus \& Diamond 1995, Gabuzda 1999, Gabuzda et al. 2004, Kharb 
et al. 2008). To explain the observational data, simplified models of the 
three-dimensional magnetic field distribution have been proposed (see, for 
example, Laing 1981, 1993, Pushkarev et al. 2005, Laing et al. 2006 ). In 
particular the polarization data suggest that the magnetic field has a 
significant transverse component in the main fration of the jet volume. In 
accordance with the "traditional" interpretation (see, e.g., Laing 1993), this 
magnetic structure can be associated with a series of shocks along the jet, 
which enhances the local transverse field. However, 
Very Long Baseline Interferometry (VLBI) observations of 
BL Lac 1803+784 (Hirabayashi et al. 1998) revealed that the transverse 
magnetic structure can be formed by a large-scale toroidal magnetic field, and 
indeed recent observations of a radio jet in the galaxy CGCG 049-033 also 
reveal a predominantly toroidal magnetic field (Bagchi et al. 
2007, Laing et al. 2006). 

Magnetic structures with a significant toroidal field-component are predicted 
by several   theoretical models for jet formation (Blandford \& Payne 1982, 
Romanova \& Lovelace 1992, Koide, Shibata \& Kudoh 1998). In particular some 
models of the jet propagation (see, e.g., Begelman, Blandford \& Rees 1984) 
propose that the toroidal field decays more slowly than the poloidal one as 
a function of the  distance from the central object, therefore the toroidal 
field should eventually become dominant.  Other generation mechanisms based 
either on the combined influence of turbulence and large-scale shear (see, 
e.g. Urpin 2006) or on dynamo action along with magneto-centrifugal and 
reconnection processes (de Gouveia Dal Pino, 2005) also predict the occurence 
of a significant  toroidal component. A situation where loops of the toroidal 
magnetic field dominate the field distribution would produce important 
dynamical consequences for the jet structure, for instance, providing 
pressure confinement through magnetic tension forces (Chan \& Henriksen 
1980, Eichler 1993). However, it is well known from plasma physics (see, e.g., 
Kadomtzev 1966) that the toroidal magnetic field may cause various magnetohydrodynamic (MHD) 
instabilities even in simple cylindrical configurations. 

In general, different types of instabilities can occur in astrophysical 
jets. They can be caused by boundary effects, shear, stratification and 
magnetic fields  and  are also likely to be responsible for the observed 
morphological complexity of jets. For this reason several analytical (see,
e.g., Ferrari, Trussoni \& Zaninetti 1980, Bodo et al. 1989, 1996, Hanasz, 
Sol \& Sauty 1999; see also Birkinshaw 1997 for a review) and numerical 
(Hardee et al. 1992, Appl 1996, Lucek \& Bell 1996, Min 1997, Kudoh, 
Matsumoto \& Shibata 1999) works have been devoted to the study of the 
stability properties of jets. A systematic study of the Kelvin-Helmholtz 
instability in non-relativistic jets with the longitudinal magnetic field 
has been performed by Bodo et al. (1989, 1996) including also the effect of rotation. 
In particular  several unstable modes including a slow 
mode associated with the magnetic field and an inertial 
mode caused by rotation have been found. Shear-driven instabilities  
can also be very important in jets (Urpin, 2002). Numerical simulations (Aloy 
et al. 1999a, 1999b) indicate that the radial structure of jets may be more 
complex with a transition shear layer surrounding the jet core. The 
thickness of the shear layer depends on the distance being around 20\% 
of the jet radius near the nozzle but broadening near the head where it is 
of the order of the jet radius. This shear layer may have an important 
consequence for various properties of jets including their stability (see, 
e.g., Hanasz \& Sol 1996, 1998, Aloy \& Mimica 2008). 

In this paper, we consider the instability associated with the topology of 
the magnetic field in jets. Magnetic fields generated by  
hydrodynamic motions can be topologically non-trivial and thus prone 
to pressure-driven or current-driven instabilities (see, e.g., Longaretti 
2008 for review). This problem is well studied in the context of experiments 
on controlled thermonuclear fusion (see, e.g., Bateman 1978) and  has been
discussed in relation to  jets by a number of authors (see, e.g., Eichler 
1993, Appl 1996, Appl et al. 2000, Lery et al. 2000). A linear stability 
analysis of supermagnetosonic jets has been performed by Appl et al. (2000) 
for different magnetic configurations including a force-free field. It has 
been shown that the current-driven instabilities can occur  
in the axial field for various profiles of the azimuthal field even if these 
components are comparable. The non-linear development of the current-driven 
instability has been analysed by Lery et al. (2000), who found that this 
instability can modify the magnetic structure of a jet redistributing the 
current density in the inner part. The current-driven instability of 
relativistic jets and plerions in toroidal magnetic fields 
has been considered by Begelman (1998), who obtained that the dominant 
instability in this case are the kink  ($m=1$) and pinch ($m=0$) modes. The 
kink mode generally dominates, thus destroying the concentric field structure. 
Apart from the current-driven instabilities, the pressure-driven ones 
caused by the thermal pressure gradient can also occur in jets. The modes, 
corresponding to these instabilities, are radially localized and characterized 
by high wavenumbers. Likely, these modes are responsible for generating MHD 
turbulence in the jet core (Kersale, Longaretti, \& Pelletier 2000). The 
pressure-driven instabilities are well studied in laboratory pinch 
configurations as well. Robinson (1971) considers the stability of the pinch 
with a large ratio of the gas and magnetic  pressure using the energy 
principle and finds that stability is possible if currents are located 
outside the main plasma column. The criterion of pressure-driven instabilities
in cylindrical geometry has been derived by Longaretti (2003). This criterion 
is related to the Alfv\'enic modes, which usually require less restricting
conditions for instability and can even be unstable if Suydam's criterion of
stability is satisfied. {If both 
destabilizing factors, current and pressure gradient are present, 
then modes of mixed nature can generally show  in a jet. An example 
of these unstable modes has been studied by Baty \& Keppens (2002), who 
considered the jet to have  equal density with the surrounding medium, separated 
by a shear layer. The magnetic field in the basic state was assumed to be 
supersonic and helical with a vanishing radial component. The linear  
stability analysis  mainly addresses  the behaviour of kink modes with 
$m = \pm 1$ where $m$ is the azimuthal wavenumber. In this  basic state, 
the thermal pressure gradient in combination with the component of
the current parallel to the magnetic field produces the energy of the 
perturbations that feed the instability. The authors calculated 
the spectrum of these modes and have found that they can be unstable. 
The unstable modes are usually oscillatory and their growth rate exhibits a 
rather peculiar double peak dependence on the axial wavevector. 
}

Note that numerical simulations often support the idea that magnetic jets 
should be unstable and can generate turbulent motions caused by instabilities. 
For instance, Lery \& Frank (2000) found that the instabilities seen 
in their jet simulations develop with a wavelength and growth time that are 
well matched by a linear stability analysis. Relativistic simulations of 
extragalactic jets by Leismann et al. (2005) and Roca-Sogorb et al. 
(2008) also clearly show that the various irregularities that develop in the 
jet flow can be interpreted as MHD instabilities. 

Our study focuses on the stability of jets with a subthermal magnetic 
field where the magnetic energy is lower than the thermal energy of 
particles. This kind of magnetic field can be generated, for instance, by the 
turbulent dynamo action or by hydrodynamic motions (stretching) in the 
process of jet propagation. We show that stability properties of jets 
with a subthermal field can be essentially different from those with a 
highly superthermal force-free magnetic field as was considered, for example, by 
Appl et al. (2000). In particular, higher azimuthal modes with $m\gg1$ can
have the highest growth rate at variance with the superthermal case.
In our jet model, both destabilizing factors - the pressure gradient 
and electric current - are presented simultaneously, and the instability
occurs under the combined influence of both these factors. Therefore, the 
instability considered has a mixed pressure- and current-driven nature.

The paper is organized as follows. Section 2 contains the derivation 
of the relevant equations, and Sect. 3 shows the numerical results. 
Conclusions are presented in Sect. 4.

\section{Basic equations}

We consider a very simplified model assuming that the jet is an infinitely
long stationary cylindrical outflow and its propagation through the ambient 
medium is modelled by a sequence of quasi-equilibrium states. For the sake of 
simplicity, it is often assumed that the velocity of plasma within a jet,
$V$, does not depend on coordinates (see, e.g., Appl 1996, Appl et al. 2000, 
Lery et al. 2000), 
and we will adopt the same assumption in our study. Instabilities of 
the magnetic configurations associated to the electric current are basically 
absolute instabilities, i.e. they grow, but do not propagate. We suppose that 
this is the case also in the rest frame of the jet, and unstable perturbations 
are therefore simply advected with the flow at the jet velocity (see, e.g.,
Appl et al. 2000). Therefore, we treat the stability properties in the rest 
frame of the jet. 

We explore the cylindrical 
coordinates ($s$, $\varphi$, $z$) with the unit vectors ($\vec{e}_{s}$, 
$\vec{e}_{\varphi}$, $\vec{e}_{z}$). The magnetic field is assumed to be 
axisymmetric with a non-vanishing $z$- and $\varphi$-component. The azimuthal 
field depends on the cylindrical radius alone, $B_{\varphi}= B_{\varphi}(s)$, 
but the axial magnetic field $B_z$ is constant in our model.

We study the behaviour of MHD modes in the incompressible limit.
This approximation is well justified for modes with the characteristic
time-scale longer than the period of sound waves and for subsonic motions 
(see, e.g., Landau \& Lifshitz 1981). Perturbations of the density caused by
these motions are small and can be neglected in MHD equations. By making use 
of the incompressible limit in a magnetised gas, one can consider slow modes 
with the growth rate (or frequency) lower than the frequency of fast 
magnetosonic waves. Since the typical time scale of modes under study is 
of the order of the inverse Alfv\'en frequency, our consideration applies 
if the Alfv\'en velocity is lower than the sound speed or, in terms of the 
plasma $\beta$-parameter, if $\beta \gg 1$ ($\beta$ is the ratio of the gas 
and magnetic pressures). Note that the velocity of jet $V$ can be much 
higher than the Alfv\'en and sound speed. In the incompressible limit, 
the MHD equations read 
\begin{eqnarray}
\frac{\partial \vec{v}}{\partial t} + (\vec{v} \cdot \nabla) \vec{v} = 
- \frac{\nabla P}{\rho} 
+ \frac{1}{4 \pi \rho} (\nabla \times \vec{B}) \times \vec{B}, 
\end{eqnarray}
\begin{equation}
\nabla \cdot \vec{v} = 0, 
\end{equation}
\begin{equation}
\frac{\partial \vec{B}}{\partial t} - \nabla \times (\vec{v} \times \vec{B}) 
= 0,
\end{equation}
\begin{equation}
\nabla \cdot \vec{B} = 0. 
\end{equation}
In the basic state, the gas is assumed to be in hydrostatic equilibrium in
the radial direction, then
\begin{equation}
\nabla P = \frac{1}{4 \pi} (\nabla \times \vec{B}) \times \vec{B} .
\end{equation}
Stability will be studied by making use of a linear perturbative analysis.
Because the basic state is stationary and axisymmetric, the dependence of 
perturbations on $t$, $\varphi$, and $z$ can be taken in the form 
$\exp{(\sigma t - i k_z z - i m \varphi)}$ where $k_z$ is the wavevector in 
the axial direction and $m$ is the azimuthal wavenumber. Small perturbations 
will be indicated by subscript 1, while unperturbed quantities will have no 
subscript. Then, the linearized Eqs.~(1)-(4) read
\begin{equation}
\sigma \vec{v}_{1} = - \frac{\nabla P_{1}}{\rho} + 
\frac{1}{4 \pi \rho}[ (\nabla \times \vec{B}_{1}) \times \vec{B} 
+ (\nabla \times \vec{B}) \times \vec{B}_{1}], 
\end{equation}
\begin{equation}
\nabla \cdot \vec{v}_{1} = 0, 
\end{equation}
\begin{equation}
\sigma \vec{B}_{1} - (\vec{B} \cdot \nabla) \vec{v}_{1} + (\vec{v}_{1}
\cdot \nabla) \vec{B} = 0,
\end{equation}
\begin{equation}
\nabla \cdot \vec{B}_{1} = 0. 
\end{equation}
Eliminating all variables in favour of the radial velocity perturbation 
$v_{1s}$, we obtain 
\begin{eqnarray}
\frac{d}{ds} \left[ \frac{1}{\lambda} (\sigma^2 + \omega_A^2) \left( 
\frac{d v_{1s}}{ds} + \frac{v_{1s}}{s} \right) \right] 
- k_z^2 (\sigma^2 + \omega_A^2) v_{1s} -  \nonumber \\
2 \omega_{B} \left[ k_z^2 \omega_{B} (1 \! - \! \alpha) \! - \!
\frac{m (1 \! + \! \lambda)}{s^2
\lambda^2} \left( 1 \! - \! \frac{\alpha \lambda}{1 \! + \! \lambda} \right) 
(\omega_{Az} \! + \! 2 m \omega_{B} ) \right.
\nonumber \\
\left. - \frac{m \omega_{Az}}{s^2 \lambda^2} \right] v_{1s} + 
\frac{4 k_z^2 \omega_{A}^2 \omega_{B}^2}{\lambda (\sigma^2 + \omega_{A}^2)} v_{1s}
=0, \;\;\;\;\;\;
\end{eqnarray} 
where
\begin{eqnarray}
\omega_{A} =  \frac{1}{\sqrt{4 \pi \rho}} \left( k_z B_z + \frac{m}{s} 
B_{\varphi} \right), \;\;
\omega_{Az} =  \frac{k_z B_z}{\sqrt{4 \pi \rho}} , \;\;
\nonumber \\
\omega_{B} \! = \! \frac{B_{\varphi}}{s \sqrt{4 \pi \rho}} , \;\;
\alpha \! = \! \frac{\partial \ln B_{\varphi}}{\partial \ln s} , \;\;
\lambda = 1 + \frac{m^2}{s^2 k_z^2} . \nonumber 
\end{eqnarray}
This equation was first derived by Bonanno \& Urpin (2008b) in their analysis
of the non-axisymmetric stability of stellar magnetic configurations. For 
axisymmetric perturbations ($m=0$), Eq.~(10) recovers Eq.~(11) of the paper 
by Bonanno \& Urpin (2008a). 

Instabilities of the magnetic configurations associated to the electric 
current are basically absolute instabilities, i.e. they grow but do not 
propagate. We suppose that this is the case also in the rest frame  of the 
jet, and unstable perturbations are therefore simply advected with the flow 
at the jet velocity. Equation (10) represents a non-linear eigenvalue problem 
for $\sigma$, which can be solved once the boundary conditions are given. 
We assume that $v_{1s}$ should be finite at the jet axis. As far as the 
outer boundary is concerned, it is somewhat difficult to formulate a plausible 
boundary condition  because actually there is no boundary between the jet 
and the ambient medium. Likely, the jet is separated from the ambient plasma by
the shear layer that has a finite thickness. The effect of this thickness
on the growth rate of instabilities has been studied by Baty (2005) who
found that the results are only slightly affected by a change of the 
thickness. For the sake of simplicity, therefore, we assume that the shear
layer is infinitely thin (see also Appl et al. 2000). In a supermagnetosonic 
jet, no signal can propagate from the jet interior to its surroundings,
and we can expect that the instabilities behave as if the jet is
bounded by a rigid conducting wall. Therefore, we can mimic  
the outer boundary by supposing $v_{1s}=0$ at the jet radius $s=s_1$.  Note 
that we also tried other outer boundary conditions (for example, with $v_{1s} 
\neq 0$), but this does not change the results qualitatively.

We can represent the azimuthal magnetic field as 
\begin{equation}
B_{\varphi} = B_{\varphi 0} \psi(s),
\end{equation} 
where $B_{\varphi 0}$ is the characteristic field strength and $\psi \sim 1$. 
To calculate the growth rate of the instability, it is  convenient to 
introduce dimensionless quantities
\begin{equation}
x \!= \! \frac{s}{s_1},\;\; q \!= \! k_z s_1,\;\; \Gamma \! = \!
\frac{\sigma}{\omega_{B0}}, \;\; \omega_{B0} \!= \! \frac{B_{\varphi 0}}{s_1
\sqrt{4 \pi \rho}}, \;\;
\varepsilon \! = \! \frac{B_z}{B_{\varphi 0}}.  
\end{equation}
Then, Eq.~(10) reads
\begin{eqnarray} \label{pert}
\frac{d}{dx} \! \left( \! \frac{d v_{1s}}{dx} \! + \! \frac{v_{1s}}{x} 
\! \right) \! + \!
\left( \! \frac{d v_{1s}}{dx} \! + \! \frac{v_{1s}}{x} \! \right) 
\frac{d \ln \Delta}{d x}
\! - \! q^2 \! \left( \! 1 \! + \! \frac{m^2}{q^2 x^2} \! \right) v_{1s} \! -
\nonumber \\
\frac{2 q^2 \psi(x)}{x (\Gamma^2 \! \! + \! \! f^2)} \left\{ \! \left[ \! 
\left( \! 1 
\! - \! \frac{m^2}{q^2 x^2} \! \right) \! \frac{\psi(x)}{x} \!-\!
\frac{m \varepsilon}{q x^2} \! \right]
(1 \! - \! \alpha) \! - \! \frac{2 m f}{m^2 \! \! + \! \! q^2 x^2} \!  \right\} 
v_{1s} 
\nonumber \\
+ \frac{4 q^2 f^2 \psi^2(x)}{x^2 (\Gamma^2 + f^2)^2} v_{1s} = 0, \;\;\;\;\;\;
\end{eqnarray}
where
\begin{equation}
f = q \varepsilon + m \frac{\psi(x)}{x}, \;\; \Delta = 
\frac{q^2 x^2 (\Gamma^2 +
f^2)}{m^2 + q^2 x^2}.  
\end{equation}
We thus solve Eq.~(13) for a wide range of the parameters and different 
functional dependence $\psi(x)$.

\section {Numerical results}

We assume that the function $\psi(s)$,  determining the dependence of 
the azimuthal magnetic field on the radial coordinate,  is given by
\begin{equation}
\psi(s)  = x^{n} e^{-p(x-1)},
\end{equation} 
where $n$ and $p$ are the parameters. The function $\psi$ goes to 0 
at the jet axis and is equal to 1 at the surface $x=1$ for any $n$ and $p$. 
For the purpose of illustration,we plot $\psi(x)$ for several values of 
$p$ and $n$ in Fig.~1. Field profile (15) with $n=1$ corresponds to the 
magnetic configurations with non-vanishing current at the jet axis, whereas 
this current goes to 0 for $n=2$. The total axial current, 
integrated over the jet section, is the same for both cases. Therefore, the
distributions with $n=1$ and $n=2$ can be the representatives of the jet 
models with the current density being relatively smoothed and concentrated 
to the outer boundary, respectively. The function (15) has the maximum at 
$x_m = n/p$ and therefore $B_{\varphi}$ reaches the maximum at the jet 
boundary if $n\geq p$ or inside the jet if $p>n$. Therefore, the cases
$p=1, n=1$ and $2$ represent the jet models with the maximum toroidal 
field at the outer boundary, whereas the field has its maximum within 
the jet for $p=2, n=1$. Because the real 
distribution of the magnetic field is unknown in jets, we will use Eq.(15) 
to mimic different possible dependences of $B_{\varphi}$ on $x$. As far 
as the axial field is concerned, we assume that $B_z$ is constant. In real 
jets, however, $B_z$ can depend on $x$ as well. Generally, this dependence 
does not qualitatively change the stability properties (see Bonanno \& 
Urpin 2008a), but it can provide an additional destabilizing effect if 
$B_z$ changes the sign inside the jets (Robinson 1971). 

\begin{figure}
\includegraphics[width=9.0cm]{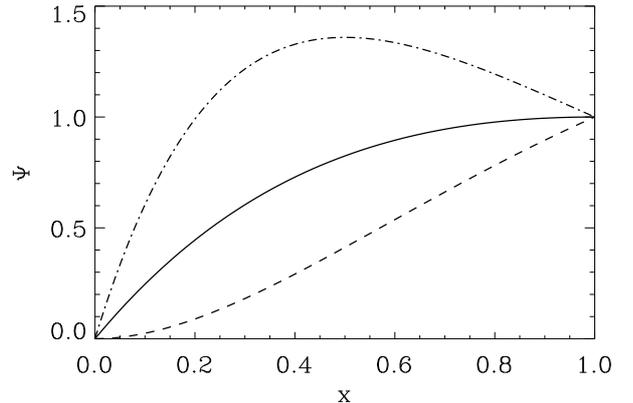}
\caption{Dependence of $\psi$ on $x$ for $p=1, n=1$ (solid line),
$p=1, n=2$ (dashed line), and $p=2, n=1$.}
\end{figure}

{ In our jet model, the electric current, $\vec{j} = (c/4 \pi) \nabla
\times \vec{B}$, only has a $z$-component, whereas $\vec{B}$ has both $z$- 
and $\varphi$-components. Therefore, the current $\vec{j}$ can be 
represented as a sum of components parallel and perpendicular to the 
magnetic field, $\vec{j}_{\parallel}$ and $\vec{j}_{\perp}$, respectively.
The parallel component does not contribute to the Lorentz force. However, even 
if $\vec{j}_{\perp}=0$ and the magnetic field is force-free, this component alone 
can produce an  unstable  magnetic configuration. 
This type of instability is usually called current-driven. On the 
contrary, $\vec{j}_{\perp}$ contributes to the Lorentz force, which should then
be balanced by a pressure force in the basic state. The magnetic 
configurations with $\vec{j}_{\perp} \neq 0$ and $\vec{j}_{\parallel} 
= 0$ can be unstable as well, and this type of instability is called 
pressure-driven. In our model, both destabilizing
factors are presented and, most likely, the instability is of the mixed
nature. To some extent, the relative importance of pressure- and 
current-driven effects can be characterized by the ratio $\vec{j}_{\perp} /
\vec{j}_{\parallel}$ in the basic state. This ratio is high if $B_z 
\ll B_{\varphi 0}$ ($\varepsilon \ll 1$) and the pressure-driven effects
determine stability properties. On the contrary, the current-driven 
effects are more important if $B_z \gg B_{\varphi 0}$ ($\varepsilon \gg 1$).   
In Fig.2 we plot the ratio $\vec{j}_{\perp} / \vec{j}_{\parallel}$ as
a function of the radius for different values of parameters determining 
the magnetic configuration. Evidently the pressure- and 
current-driven effects generally are comparable if $\varepsilon$ is of 
the order of 1.     
}  

\begin{figure}
\includegraphics[width=9.0cm]{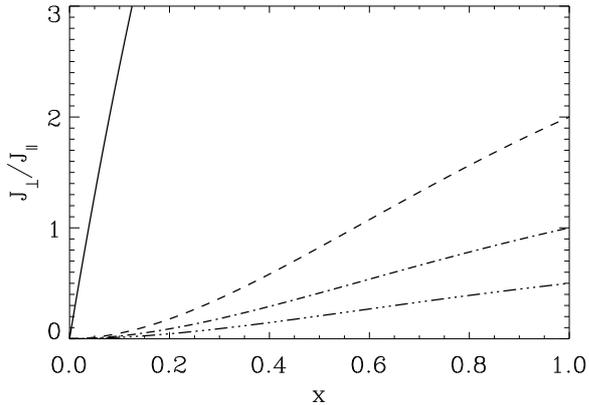}
\caption{Ratio of perpendicular and parallel to the magnetic field 
components of the electic current on $x$ for $\varepsilon = 0.1$, $p=1, 
n=1$ (solid curve), $\varepsilon = 0.5$, $p=1, n=2$ (dashed),
$\varepsilon = 1$, $p=1, n=2$ (dot-dashed), and
$\varepsilon = 2$, $p=1, n=2$ (dot-dot-dashed).}
\end{figure}

Equation~(13) together with the given boundary conditions is   
a two-point boundary value problem, which can be solved by using the ``shooting''
method (Press et al. 1992). In order to solve Eq.~(13), we used a 
fifth-order Runge-Kutta integrator embedded in a globally convergent 
Newton-Rawson iterator. We  checked that the eigenvalue was always the 
fundamental mode because the corresponding eigenfunction had no zero 
except for that at the boundaries. 

In Fig.~3 we plot the dependence of the growth rate on the axial wavevector
for the magnetic configuration with non-vanishing current at the axis and
a relatively weak axial magnetic field, $\varepsilon = B_z/ B_{\varphi 0}=
0.1$. As  was noted by Bonanno \& Urpin (2008),  an axial   
field breaks the symmetry between positive and negative values of the axial
wavevector $q$ in Eq.~(13) even if $B_z$ is small compared to $B_{\varphi}$. 
However, Eq.~(13) still contains some degeneracy because it is invariant under
$(m, q) \rightarrow (-m, -q)$ or $(m, \varepsilon) \rightarrow (-m, 
-\varepsilon)$ transformation. The instability occurs only for 
perturbations with $q$ within a narrow range that depends on the 
azimuthal wavenumber $m$. For example, perturbations with $m=1$ and $m=6$ 
are unstable if $0 > q > -70$  and $-60 > q > -200$, respectively. Note 
that the unstable perturbations with positive $m$ should have negative $q$ 
and, on the contrary, if $m$ is negative, instability occurs only for 
perturbations with positive $q$.

The growth rate has two clear maxima with the highest maximum 
corresponding to $q \sim - m/ \varepsilon$. By the order of magnitude, the 
axial wave-vector of the most rapidly growing perturbation can be estimated 
from the condition of magnetic resonance
\begin{equation}
\omega_A = \frac{1}{\sqrt{4 \pi \rho}} \left( k_z B_z + \frac{m}{s} 
B_{\varphi} \right) \approx 0.
\end{equation}     
Indeed, this equation implies $\omega_A \propto q 
\varepsilon + m \psi(x) \approx 0$. Since $\psi(x) \sim 1$ in our 
model, the condition $\omega_A = 0$ corresponds to 
\begin{equation}
q \sim - m /\varepsilon.
\end{equation} 
The most rapidly growing modes turn out to be highly anisotropic if the 
axial field  is weak compared to the toroidal one: their axial wavelength 
$\lambda_z = 2 \pi/k_z \sim 2 \pi \varepsilon s$ is much shorter than 
the radial and azimuthal lengthscale. The growth rate is fairly high and 
 is of the order of the inverse Alfven time scale. 
The growth rate slowly increases with $m$ and  perturbations with 
a shorter azimuthal scale grow faster. The axisymmetric mode ($m=0$) turns 
out to be the most slowly growing.

\begin{figure}
\includegraphics[width=9.0cm]{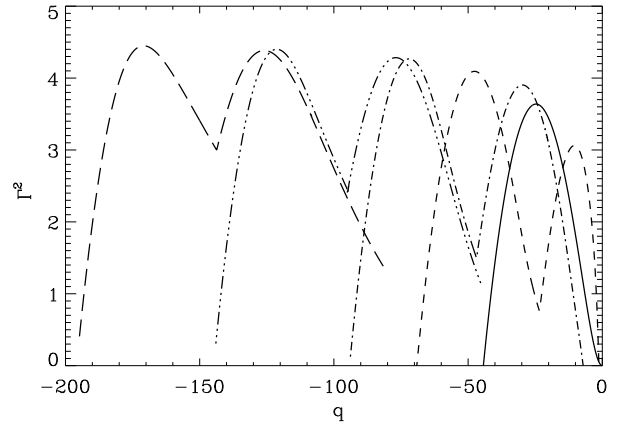}
\caption{Dependence of the normalized growth rate of instability 
on the wavevector $q$ for $\varepsilon=0.1$, $p=1$ and $n=1$. Different types
of lines correspond to different azimuthal wavenumbers: $m=0$ (solid), $m=1$
(dashed), $m=2$ (dot-and-dashed), $m=4$ (dot-dot-dashed), and $m=6$ 
(long-dashed).}
\end{figure}

The model with $p=1, n=2$ exhibits a similar behaviour of perturbations (see
Fig.~4). As mentioned, electric currents are more concentrated near the outer 
boundary in this model. The values of $q$ that allow the instability are smaller 
in this case and therefore the corresponding vertical wavelengths are longer, but 
the range of unstable $q$ is  narrower. For the considered values of $m$,
the growth rate is lower approximately by a factor 2. The general impression 
is that the configuration with currents concentrated closer to the outer
boundary is more stable than that with more uniformly distributed currents.
This conclusion qualitatively agrees with the result obtained
by Robinson (1971) from the hydromagnetic energy principle. The author considers
a pinch configuration with large $\beta$ and finds that the configuration is
more stable if large axial currents flow outside the main plasma column.

Note that the equilibrium state of both the configurations considered in 
Fig.~3 and 4 is characterized by the negative pressure gradient that is required
for the development of instability (Longaretti 2008). Indeed, using Eq.~(5) and
expression (15) for the toroidal field, we obtain
\begin{equation}
\frac{d P}{d s} = - \frac{B_{\varphi}^2}{4 \pi s_1} \left( \frac{n+1}{x} - 
p \right).
\end{equation}
Evidently $d P/d s < 0$ everywhere within the range $1 > x > 0$ if $p=1$ 
and $n=1,2$. The sign of the pressure gradient is important because it determines
the destabilizing effect in the so called Suydam's criterion. This criterion 
represents a necessary condition for stability (see, e.g., Longaretti 2003) and
reads in our notations 
\begin{equation}
\frac{s B_z^2}{4 \pi} \left( \frac{1}{h} \frac{d h}{d s} \right)^2 
+ 8 \frac{d P}{d s} > 0 ,
\end{equation}
where $h = s B_z / B_{\varphi}$ is the magnetic shear. For the
equilibrium configuration with the toroidal field given by Eq.~(15), this
criterion can be rewritten as
\begin{equation}
\frac{\varepsilon^2}{8 x} ( 1 - n + px)^2 - \psi^2 \left( \frac{n+1}{x}
- p \right) > 0 .
\end{equation}
For the chosen parameters ($p=1, n=1,2$), the necessary condition for stability
is not satisfied in some fraction of the jet volume (for example, near the 
outer boundary), and the corresponding configurations can generally be 
unstable.

\begin{figure}
\includegraphics[width=9.0cm]{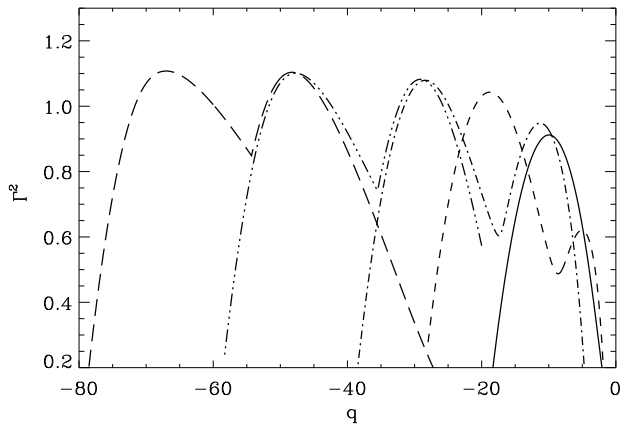}
\caption{Dependence of the normalized growth rate of instability 
on the wavevector $q$ for $\varepsilon=0.1$, $p=1$ and $n=2$. Different 
types of lines correspond to the same azimuthal wavenumbers as in Fig.2.}
\end{figure}

Figure~5 shows the dependence of the growth rate on $q$ for the model
with $p=2, n=1$. In this case,  the maximum of the azimuthal field is located 
within the jet volume, at $x=0.5$.
The maximum value is approximately 1.3 times higher than the boundary 
one. It appears that the instability grows  slightly faster in this type of
magnetic configurations, but all the main qualitative features of the
instability are unchanged. Because $B_{\varphi}$ is stronger in this
model, the instability occurs for larger $q$ as follows from Eq.~(16).
Note that we calculate $\Gamma$ only for $m \leq 3$ because of computational
problems at larger $m$.


\begin{figure}
\includegraphics[width=9.0cm]{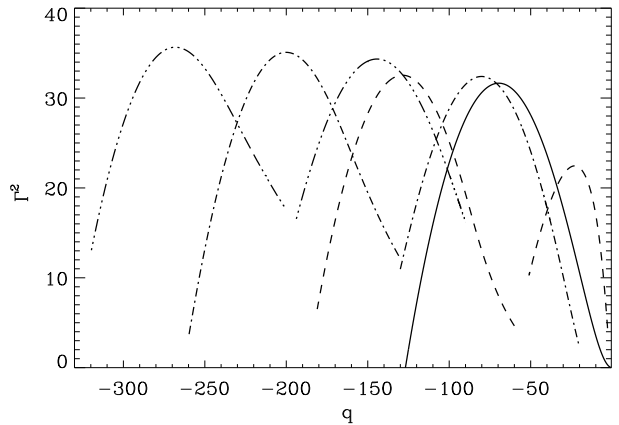}
\caption{Dependence of the normalized growth rate of instability 
on the wavevector $q$ for $\varepsilon=0.1$, $p=2$ and $n=1$. Different 
types of lines correspond to different $m$: $m=0$ (solid), $m=1$ (dashed),
m=2 (dot-and-dashed), and $m=3$ (dot-dot-dashed). }
\end{figure}

In Fig.6 we plot the growth rate of instability for the configuration with 
the axial field much weaker than the toroidal one, $\varepsilon = 0.01$. 
The distribution of the toroidal field is given by Eq.~(15) with $p=1$ and 
$n=2$. The instability can occur in this case as well, but only perturbations
with very large $q$ can be unstable. Even for the mode $m=1$, the growth 
rate reaches its maximum at $q \approx 200$. For modes with larger $m$, the
maximum growth rate corresponds to substantially higher values of $q$ (very
short axial wavelengths). Unfortunately, our numerical code does not allow us
to resolve the modes with a very short axial wavelength, and we were able
to extend our calculations only to $q \approx 300$. Note that  
that even perturbations with a $q$ this high still grow rapidly and the 
growth rate is of the order of the inverse Alfven time.    

\begin{figure}
\includegraphics[width=9.0cm]{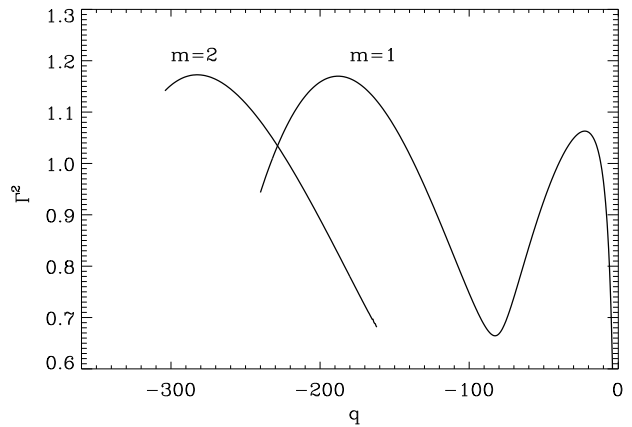}
\caption{Growth rate as a function of $q$ for $\varepsilon = 0.01$
and distribution of the toroidal field (15) with $p=1$ and $n=2$.}
\end{figure}

Figure 7 shows the growth rate versus $q$ for a higher ratio of the axial
and toroidal fields ($\varepsilon=0.5$) and for the model with $p=1$
and $n=2$. In this case, the instability occurs at substantially lower
values of $q$ and, correspondingly, at longer axial wavelength. The modes
with small $m$ grow relatively slowly, but the growth rate increases with 
increasing $m$. The mode with $m=8$ has the growth rate comparable
to the inverse Alfven time, but the modes with larger $m$ are growing even 
faster. In that sense, the dependence of modes on $m$ in the cases 
$\varepsilon \sim 1$ and $\varepsilon < 1$ seems to be qualitatively 
similar.    

\begin{figure}
\includegraphics[width=9.0cm]{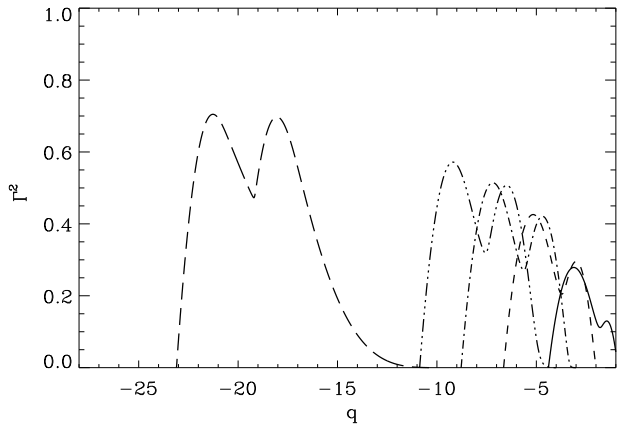}
\caption{Growth rate as a function of $q$ for $\varepsilon = 0.5$,
$p=1$ and $n=2$. The curves correspond to $m=1$ (solid), $m=2$ (short-dashed),
$m=3$ (dot-and-dashed), $m=4$ (dot-dot-dashed) and $m=8$ (long-dashed).}
\end{figure}

Figure 8 compares the growth rates for the cases $\varepsilon=1$ and $2$. We
considered the model with $p=1$ and $n=2$, but the behaviour is 
qualitatively the same as for the other models. The growth rates are generally 
lower than in the previous cases ($\varepsilon<1$), but the instability 
is still present. A suppression of the growth rate is certainly more
pronounced if $\varepsilon =2$. However, even in this case, $\Gamma$ increases 
relatively rapidly with increasing $m$ and it is likely to  reach a saturation 
value $\sim 1$ at sufficiently large $m$. The values of $q$ that correspond 
to unstable perturbations are lower than in Figs.3-6. Figure 8 illustrates very 
well  that the instability cannot be completely suppressed  even if 
the axial field is sufficiently strong and $\varepsilon > 1$. At large 
$\varepsilon$, the instability 
is not very efficient ($\Gamma^2>0$ but very small)  for perturbations 
with not very large $m$ but those with large $m$ can 
still reach a saturation value of $\Gamma^2 \sim 0.1$.  
Indeed in the specific case of $\varepsilon=2$ we could not find 
significant growth rate for modes with $m \leq 5$, but 
we found that the growth rate is of the order of the inverse Alfven crossing 
time for sufficiently high azimuthal wavenumbers.
     
\begin{figure}
\includegraphics[width=9.0cm]{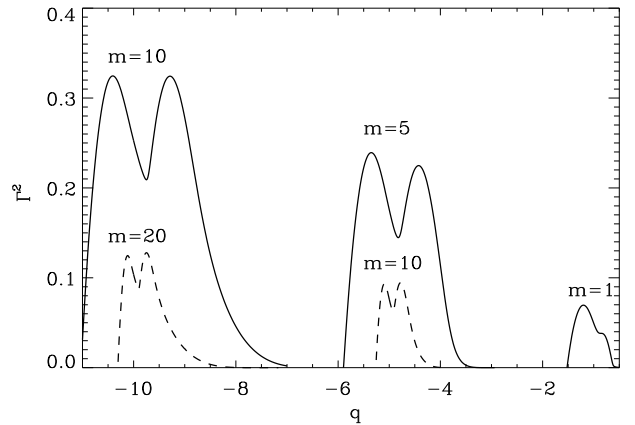}
\caption{Comparison of the growth rate for the cases $\varepsilon=1$ (solid)
and $2$ (dashed).The parameters determining the distribution of the toroidal
field are $p=1$ and $n=2$.}
\end{figure}

To illustrate how velocity perturbations depend on the cylindrical radius,
in Fig.~9 the eigenfunctions for one of the models shown in Fig.~8 
($\varepsilon = 1$) are plotted. The eigenfunctions correspond to the azimuthal 
wavenumbers
$m=1$ (solid), $5$ (dashed), and $10$ (dot-and-dashed). The $m=1$ mode is the only
mode that does not go to zero at the jet axis. Typically, the modes reach their 
maximum closer to the outer boundary, and an increase of $m$ results in the 
eigenfunctions that are more and more localized. For large $m$, motions in
the core region of the jet are substantially suppressed. This sharpness of the
eigenfunctions is the main reason of the problems in numerical calculations for
high $m$. 

\begin{figure}
\includegraphics[width=9.0cm]{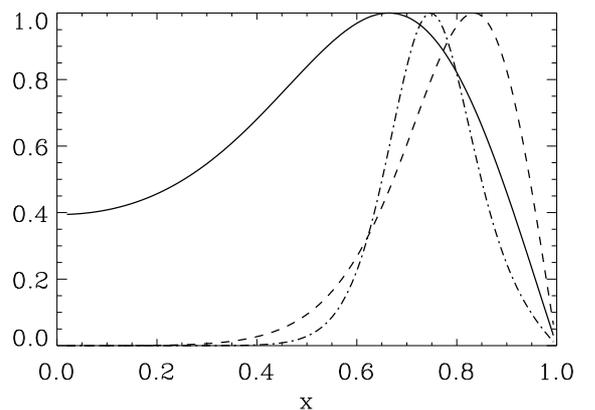}
\caption{The eigenfunctions for the model with $\varepsilon = 1$,
$p=1$, and $n=2$. The curves correspomd to $m=1$ (solid), $m=5$
(dashed), and $m=10$ (dot-and-dashed).}
\end{figure}

\section{Discussion}

We considered the instability that can occur in jets if the magnetic 
field has a relatively complex structure with non-vanishing azimuthal and 
axial components. A magnetic field with complex topology 
is suggested by both  the theoretical models for jet formation and the 
mechanisms of generation of the magnetic field. Our study considers only 
jets with subthermal magnetic fields because the the stability properties of 
jets with superthermal fields can be essentially different (Appl et al. 
2000).

The considered model is very simplified and does not take into 
account many physical factors that can be important in the evolution of 
real jets. For instance, we model the transition from the jet to the
external medium in terms of the infinitely thin boundary layer. In fact,
the boundary can be somewhat extended and the velocity can change smoothly 
within the transition layer. Shear has  a destabilising 
influence on jets (Urpin 2002), and the shear-driven instability can often
be faster than instabilities associated to electric currents. An interplay 
of the shear-driven effects with the magnetic instabilities will be
considered elsewhere. Another important assumption concerns the topology
of the background magnetic field. Generally, the magnetic geometry of jets 
can be much more complicated than that considered in our paper and can 
include magnetic structures of various scales. Likely, the influence of 
these structures on the stability of jets can be studied only by using 3D 
numerical modelling. One more essential assumption of our model concerns
the lack of hydrodynamical motions in the basic state of the jet in a 
co-moving frame. The presence of motions with the velocity comparable to 
the sound speed (for example, rotation) can drastically change the 
stability properties.

It turns out that the magnetic jet is always unstable if the azimuthal 
field increases to the outer boundary. This does not depend on the 
ratio of the axial and azimuthal fields. However, the character of the 
instability can be essentially different for different $\varepsilon$. If 
the axial field is weaker than the azimuthal one and $\varepsilon < 1$, 
the growth rate tends to a saturation  value of the order of the 
inverse of a few Alfven crossing times for $m>1$. Our code does not allow 
us to calculate eigenfunctions and eigenvalues for very large $m$ in this 
case, but Figs. 3 and 4 indicate that the growth rate shows a saturation 
trend for  $m \gg 1$. Then the fastest growing mode has the axial 
wavevector $k_z \sim  - m/ \varepsilon s$,  which is determined by the 
condition of magnetic resonance. This wavevector corresponds to the 
axial wavelength $\lambda_z =2 \pi / |k_z| \sim (2 \pi \varepsilon)(s/m)$. 
At small $\varepsilon$, the axial wavelength of the most rapidly growing 
perturbations should be very small. Therefore, if the longitudinal 
magnetic field is weaker than the azimuthal one, we expect that the 
instability generates structures in jets with a very short length scale in 
the $z$-direction. Note that this behaviour can cause problems in numerical 
modelling of the instability because a very high resolution in the axial 
direction is required. 

When the ratio of the axial and azimuthal fields becomes comparable to or 
greater than 1, the behaviour of the instability as a function of $m$ is 
qualitatively similar, although the growth rate is significantly smaller. 
In this case perturbations with relatively small $m$ are substantially 
suppressed, but we still find a numerical evidence for a saturation trend 
as we increase $m$, although it is reached for much higher values of $m$ 
than in the $\varepsilon< 1$ case. It is important to stress that even $m 
\sim 100$ in the azimuthal direction is still a macroscopic scale, and 
dissipative effects are negligible in this case. 

The conlusion that the instability can arise even if $\varepsilon 
\geq 1$ with a characteristic azimuthal wavenumber $m\gg 1$ is at variance 
with the  widely accepted opinion that magnetic configurations should be 
stabilised at $B_z \sim B_{\varphi}$. Note that the similar conclusion
has been obtained by Goedbloed \& Hagebruk {1972) for the magnetic
configuration with the constant pitch, $B_{\varphi} / s B_z=const$.

{ The dependence of the growth rate on $q$ has generally a very 
particular shape with two maxima. The origin of these maxima in $\Gamma^2$
is caused by the resonant nature of unstable modes. The resonance is 
associated with the last term on the l.h.s. of Eq.(13) (or Eq.(10)) 
which is proportional to $f^2/(\Gamma^2 + f^2)^2$. For the sake of simplicity,
we can consider the case $p=0$, $n=1$. Then, $\psi / x =1$ and the 
function $f$ (Eq.(14)) characterizes a departure of $q$ from the resonant 
value given by Eq.(17). This resonant dependence reaches its maximum 
value (equal to $1/ 2 \Gamma^2$) at $f^2 = \Gamma^2$. At given $\Gamma$, 
this condition can be fulfilled for two values of $q$,
\begin{equation}
q = - \frac{m}{\varepsilon} \pm \frac{\Gamma}{\varepsilon} .
\end{equation} 
These two values of $q$ correspond to two maxima in $\Gamma^2(q)$. 
Note that the gap between the maxima $\approx 2 \Gamma
/ \varepsilon$ increases with decreasing $\varepsilon$ and does not 
depend on $m$,
which agrees qualitatively and quantitatively well with
the results of our numerical calculations.}

{ Note that the detailed physical nature and many properties of the 
instability investigated in our paper are not well understood yet.}
Our study also does not allow us to answer the question what type
of the magnetic configuration is formed because of the development of 
instability. To answer this question, one needs 3D numerical simulations. 
However, it is possible that the much better understanding of the nature
of the magnetic field in jets is required to answer this question. 
For example, the resulting large scale magnetic field can be formed
by balancing the rate of field decay caused by instability and the rate
of generation owing to some mechanism (e.g., dynamo). It would be  
important to investigate the consequences of our findings for jets by 
performing more realistic numerical simulations, and we hope to address 
this question in the near future.

\section*{Acknowledgement}

VU thanks the INAF-Osservatorio Astrofisico di Catania 
for hospitality and financial support.

\end{document}